\documentclass[12pt]{article}
\newcommand{\fr}{\frac}
\newcommand{\lb}{\label}
\newcommand{\ti}{\tilde}
\newcommand{\be}{\begin{equation}}
\newcommand{\ee}{\end{equation}}
\newcommand{\ba}{\begin{array}}
\newcommand{\ea}{\end{array}}
\newcommand{\beqa}{\begin{eqnarray}}
\newcommand{\al}{\alpha}
\newcommand{\ka}{\kappa}

\newcommand{\La}{\Lambda}
\newcommand{\si}{\sigma}

\newcommand{\del}{\partial}
\newcommand{\eeqa}{\end{eqnarray}}
\newcommand{\ep}{\epsilon}

\newcommand{\Vc}{{\cal V}}
\newcommand{\cN}{{\cal N}}

\newcommand{\kd}{\delta}

\newcommand{\bi}{\bibitem}

\title{}
\author{}
\date{}

\begin{document}

\noindent
{\bf \large The relations between 
generalized fields and superfields
formalisms  of the Batalin--Vilkovisky  method of 
\mbox{quantization} }

\vspace{2cm}

\noindent
\"{O}mer F. DAYI \\ 
{\small The Abdus Salam ICTP, 
Strada Costiera 11, 34014, Trieste--Italy \\
 Physics Department, Faculty of Science and
Letters, Istanbul Technical University,
 80626 Maslak--Istanbul,
Turkey\\
 Feza G\"{u}rsey Institute,
 P.O.Box 6, 81220
\c{C}engelk\"{o}y--Istanbul, Turkey \\
E-mail: dayi@itu.edu.tr,  dayi@gursey.gov.tr }

\vspace{2cm}

{\footnotesize
\noindent
A general solution of
the Batalin--Vilkovisky master
equation was formulated 
 in terms of generalized fields.
Recently, a superfields
approach of obtaining solutions of the  
Batalin--Vilkovisky master equation is also established.
Superfields  formalism  is  usually applied to 
topological quantum field
theories.
However, generalized fields method is suitable to
find solutions of the Batalin--Vilkovisky 
master equation either for 
topological quantum field theories or 
the usual  gauge theories like Yang--Mills theory. 
We show that
by  truncating some components of
superfields with  appropriate actions,
 generalized fields formalism
of the usual gauge theories result.
We demonstrate that
for some topological quantum 
field theories and the relativistic particle 
both of the methods possess
the same field contents and yield similar results.
Inspired by the observed relations 
we give the solution of the BV--master equation for 
on--shell $N=1$ supersymmetric Yang--Mills
theory utilizing superfields. }

\newpage

\section{Introduction}

The Batalin--Vilkovisky (BV) method offers a systematic
procedure  of finding 
actions which can be used in related path integrals
respecting symmetries like Lorentz invariance of
classical field theories possessing gauge invariance which may
be reducible\cite{bv}.
Some ad hoc definitions of the BV method of quantization 
were derived  analytically introducing an odd time 
formulation\cite{odt} which  is recently utilized to formulate some
aspects of BV method on a geometrical setting\cite{res}(for 
another approach see \cite{mv}). 
Odd time formalism 
inspired the generalized fields
 method of solving the BV--master 
equation\cite{gf1}. In this approach
one begins with a gauge theory
whose action can be written as first order in derivatives.
Although this seems to restrict the applicability
of the method drastically, it was shown that
BV--quantization of
a vast class of gauge theories can be 
obtained by this method\cite{gf1},\cite{gf2}. 
Exceptions are theories like the relativistic
superparticle where kinetic terms of anticommuting 
variables include at least  three variables.
After studying its minimal ghost content and 
antifields one introduces generalized fields
which are defined utilizing differential
form degree and BRST grading.
Substitution of original fields with generalized ones
in the original gauge theory action
yields the desired solution of the BV--master 
equation. A similar approach was also
given in \cite{bau}.

The BV method of quantization is also studied introducing 
superfields\cite{aksz},  to reveal its  geometrical
aspects\cite{park}. This method is
usually applied to find BV--master actions 
(solutions of the BV--master equation) of
topological quantum field theories.
 Superfield
algorithm is used to discuss  general solutions of
the BV--master equation in \cite{bm}. 
In terms of this method 
first order systems\cite{es} 
and  deformations of some gauge 
theories\cite{ike} are also studied 
(for another 
superfield approach see
\cite{mal} and the references given therein).
In superfields  formalism 
instead of specifying a
 gauge invariant classical 
action one starts with  an action
which can be used as a BV--master action. Underlying
gauge invariant classical action can be deduced 
by  setting ghost fields and antifields equal to zero.

Generalized fields formalism as well as
superfields method
yield  general solutions of the BV--master equation.
Purpose of this work is to discover 
the relations between these approaches.
We show that  superfields method leads 
to generalized fields solutions of the BV--master
equation  of the usual 
gauge theories like Yang--Mills theory
if some components
of superfields
are truncated in a consistent manner.
Different truncations with appropriate actions
yield different gauge theories.
However, when one deals with 
topological quantum field theories 
without any truncation, the superfields contents
coincide with generalized fields components.
This relation is not surprising: Actions
of the both methods possess the same form.
The difference can only be in field contents.
Indeed, one gets rid of this difference by 
the consistent truncations.
Understanding these relations aggregates
powers of generalized fields method and
superfields approach. This may shed light 
on understanding of different aspects of gauge theories.
Moreover, we show that 
once these relations  are discovered  they 
inspire derivation of
 BV--master actions of
some other theories, like 
supersymmetric Yang--Mills theory, in terms
of  superfields method. Here we discuss some cases which
are useful to illustrate these  relations although
there are  many other  gauge theories which can be studied
in terms of both methods. 

In the next two sections we briefly review generalized 
fields and superfields formulations of solving 
the BV--master equation to obtain  BV--master actions.
Then we show how truncations of superfields result in
generalized fields which lead to the BV--master actions
of Yang--Mills theory and the self interacting
antisymmetric tensor field  in 4 dimensions. 
Truncation of superfields are also 
 shown to be applicable to spinor fields and thus to
on--shell $N=1$ 
supersymmetric Yang--Mills theory in 4 dimensions. 
The relations between these approaches 
in other dimensions is discussed briefly,
considering the relativistic particle, Yang--Mills theory
in 2 dimensions and a topological quantum field
theory in 5 dimensions. Some other theories which
we can apply both methods are also mentioned.
In the last section we discuss the results obtained.

\section{Generalized fields method}

To obtain  BV formalism of a gauge theory, one 
introduces ghost and ghost of ghost fields if necessary
inspecting  properties of gauge transformations. 
Then, an antifield is 
assigned to each field\cite{bv}.
These fields  can
be grouped together by extending the
exterior derivative $d$  to include the  BRST transformation
$\kd_B$ as\cite{cc} 
\be
\lb{dtil}
\ti{d} \equiv d + \delta_B ,
\ee
which is defined to be nilpotent: $\ti{d}^2=0.$
Thus, we can gather the
differential form degree
$N_d$ and the ghost number $N_g$ as
the total degree
\be
\lb{tde}
\cN \equiv N_d +N_g.
\ee

We deal with  systems
whose  actions can be put into the first order form,
\be
\lb{ola}
S_0(A,B)= BdA + V(A,B)
\ee
and invariant under the gauge transformations
\be
\lb{caz}
\kd^{(0)} (A,B) = R^{(0)} (A,B)\La ,
\ee
where $\La$ is gauge parameter.
We suppress 
integration over space--time variables
and all indices.

The initial fields,  minimal set of ghost fields and
their  antifields
can be used to define generalized fields
by grouping them  in terms of the general grading
given by $\ti{d}$ as
$\ti{A}$ and $\ti{B}$ satisfying
\be
\lb{cto}
\cN (\ti{A}) = N_d (A)\ ;\ \cN (\ti{B}) = N_d (B).
\ee
Now,
substitute the original fields
$A$ and $B$ with the generalized ones
$\ti{A}$ and $  \ti{B}  $ in the action (\ref{ola}):
\be
\lb{mass}
S \equiv S_0(\ti{A}, \ti{B})= \ti{B}d\ti{A} +V(\ti{A},\ti{B} ).
\ee
Multiplication is defined such that $S$ is a
zero ghost number scalar functional.
The action
(\ref{mass}) is invariant under the transformations
\be
\lb{nonso}
\kd_{\ti{\La}} (\ti{A},\ti{B}) = \ti{R} \ti{\La},
\ee
where the generators are
\be
\lb{ggg}
\ti{R}\equiv R^{(0)}(\ti{A} , \ti{B} )
\ee
and $\ti{\La}$ is an appropriate generalization of the
original gauge parameter $\La .$
If (\ref{nonso}) can be
written as
\beqa
\left(
\begin{array}{c}
\kd_{\ti{\La}} \ti{A}_i \\
\kd_{\ti{\La}} \ti{B}_i
\end{array}
\right)   & = &
\left(
\begin{array}{cc}
-\fr{\kd_l\kd_rS}{\kd \ti{B}_i \kd \ti{A}_j} &
- \fr{\kd_l\kd_rS}{\kd \ti{B}_i \kd \ti{B}_j} \\
\fr{\kd_l\kd_rS}{\kd \ti{A}_i \kd \ti{A}_j} &
\fr{\kd_l\kd_rS}{\kd \ti{A}_i \kd \ti{B}_j}
\end{array}
\right)
\left(
\begin{array}{c}
\ti{\La}_1^j  \\
\ti{\La}_2^j
\end{array}
\right) ,
\lb{ggi}
\eeqa
where $\kd_r$ and $\kd_l$ denote 
right and left functional derivatives,
$S$ given by (\ref{mass}) satisfies
\[
(S,S)=2\fr{\kd_r S}{\kd \ti{B}_i }
\fr{\kd_l S}{\kd \ti{A}_i }=k
\]
where $k$ is a constant. We deal only with
$k=0,$ otherwise it leads to non--consistency of equations
of motion.

Here we consider the theories which can be written
in the form
\be
\lb{ef}
S =\ti{B} d\ti{A}  +\al \ti{B}
\ti{B}  +\beta \ti{A}  \ti{A}  + \gamma \ti{A} \ti{A} \ti{B} ,
\ee
where either
$\al =0$ or $\beta =0$ and the other constant
  $\gamma$  is dictated by
the original theory.

The transformations
\be
\lb{tr}
\delta_B \ti{A}  =\fr{\kd_l S}{\kd \ti{B} },\
\delta_B \ti{B}  =-\fr{\kd_l S}{\kd \ti{A} },
\ee
can be written in terms of
the covariant derivative
$\ti{D} = d+\ti{A} $ and the related curvature $\ti{F}$
when $\beta =0$ as
\be
\lb{bz}
\delta_B \ti{A}  = \ti{F}  -\ti{B}  ,\  
\delta_B \ti{B}  = -\ti{D} \ti{B}  
\ee
and when $\alpha =0$ as
\be
\lb{az}
\delta_B \ti{A}  = \ti{F}   ,\   
\delta_B \ti{B}  = -\ti{D} \ti{B}  +\ti{A}  .
\ee
The components of the right hand sides are restricted to
possess the same form degree and one more ghost number of the
components of the left hand sides.
If these transformations are nilpotent:
\be
\lb{cme}
\delta^2_B \ti{A}  =0,\   \delta^2_B \ti{B}  =0 ,
\ee
one can conclude that the BV--master equation is satisfied:
\be
(S,S)=0
\ee
In both of
the cases  (\ref{bz}), (\ref{az}) it is shown that  
(\ref{cme}) are satisfied\cite{gf1} due to 
the Bianchi identities $\ti{D} \cdot  \ti{F} =0$ and 
the definition of the curvature $\ti{F} =\ti{D} \cdot  \ti{D} .$
When  $\ti{A} =\ti{B} $ (Chern-Simons type) we have
$\alpha =0$ and $\beta =0$ in (\ref{ef}) and $\delta_B \ti{A} =\ti{F} $,
so that $\delta^2_B \ti{A} =0$ follows from the Bianchi identities.

This method is applied to some usual
gauge theories\cite{gf1},
shown to be applicable to a generalized version of Chern--Simons
theory  and BRST field theory\cite{gf2}. It is  reviewed in
\cite{gfr} as a pedagogical approach to BV--method. Moreover,
it gives an efficient  formulation of consistent 
interactions\cite{gfc}.

\section{Superfields method}

Let us deal with a superspace with $n$ commuting and
$n$ anticommuting variables: $x_\mu ,\ \tau_\mu;\ \mu =1,2,\cdots ,n,$
and the action
\be
\lb{sma}
\Sigma[U^{(n)},V^{(n)}]=\int\ d^nx\ d^n\tau \ \left(
(-1)^{\ep_U}V^{(n)}(x,\tau )D U^{(n)}(x,\tau ) - \Vc (U^{(n)},V^{(n)})
\right) .
\ee
Summation index  is suppressed.
The fields 
$V$ and $U$ possess, respectively,
Grassmann parity $\ep_U +1 $ and
$\ep_U  $ in even dimensions
and the same Grassmann parity
$\ep_U$
for odd dimensions\cite{bm}.
$D$ is the Grassmann odd, nilpotent differential operator
\be
\lb{d}
D=\tau^\mu \frac{\del}{\del x^\mu} .
\ee
Antibracket of arbitrary superfields $X,Y$ is defined as
\[
(X,Y)=
\frac{\kd_r X}{\kd V^{(n)}} 
\frac{\kd_l Y}{\kd U^{(n)}} 
-\frac{\kd_r X}{\kd U^{(n)}} 
\frac{\kd_l Y}{\kd V^{(n)}} .
\]
Boundary 
conditions and $\Vc$ are chosen such that 
the BV--master equation
is satisfied 
\be
(\Sigma ,\Sigma ) =0 
\ee
at the classical level. 

$\tau$ dependence of 
any field $X_A$  can  explicitly be written 
in terms of the components 
$X_{k}^{\mu_1\cdots\mu_k} (x)$
as
\be
\lb{ext}
X(x,\tau )=\sum_{k=0}^n \tau_{\mu_1} \cdots \tau_{\mu_k} 
X_{k}^{\mu_1\cdots\mu_k} (x).
\ee 
The unique non-vanishing integral on Grassmann variables
is normalized as
\[
\int d^n\tau \tau^{\mu_1}\cdots\tau^{\mu_n}=
\ep^{\mu_1\cdots \mu_n} ,
\]
in terms of the totally
antisymmetric tensor in n--dimensions 
$\ep^{\mu_1\cdots \mu_n}.$

Grassmann odd 
coordinates $\tau^\mu$ are defined to possess
ghost number:
\be
N_g (\tau_\mu )=1 .
\ee
One demands that 
\be
N_g (\Sigma )=0.
\ee
Thus, the superfields $U^{(n)},\ V^{(n)}$
should satisfy
\be
N_g (U^{(n)} ) + N_g (V^{(n)})=n-1.
\ee
Obviously, the components (\ref{ext}) possess
\be
N_g (X_{k}^{\mu_1\cdots\mu_k}) =N_g(X)-k .
\ee

Without implementing any restriction 
one can deal with the cases
\be
\lb{be}
N_g(V) \geq N_g(U).
\ee
Classical
gauge theory which leads to the BV--master action
$\Sigma$ can be derived from it by setting to zero the fields 
possessing nonzero ghost number.
Components of a
superfield which is defined to have negative ghost number
cannot contain vanishing ghost number. Their components
related to the underlying gauge theory can only
be antifields of some Lagrange multipliers 
of the underlying classical action.
Thus,
as far as we do not deal with 
gauge theories containing
Lagrange multipliers it is sufficient to
 consider only positive 
ghost number superfields.

\section{The relations between generalized fields and superfields 
\mbox{methods} }

The action in terms of generalized fields (\ref{mass})
as well as
 the one written by
superfields (\ref{sma}) are first
order in space--time derivatives and both of
them are  defined
to satisfy the BV--master equation.
However, their field contents can be different.
We will illustrate the relations between 
these formalisms
by focusing on some examples which are chosen 
because they represent some basic gauge systems
treated by these methods. 

(\ref{sma}) is mainly
used in topological quantum field
theories for which
it seems that in general field contents of
the both formalisms  coincide,
although we demonstrate it for some specific
cases.
However, when we consider
the usual gauge systems like Yang--Mills theory,
their field contents in general  do not coincide.
Nevertheless, 
by truncating some components of superfields consistently,
 such that
remaining ones still lead to a proper solution of 
the BV--master equation,
one obtains the generalized fields. Truncation is
performed by keeping some positive ghost number components
of fields and setting the others equal to zero. 
Obviously consistent truncations depend on the chosen
$\Vc .$ This may give the impression that a
general receipt for consistent truncations 
is missing. However, this is not the case:
Superfields contain all possible ghost number
fields available in the dimension which one
considers. Moreover, action of the superfield formalism 
is appropriately chosen which  coincides with
the form of action of the generalized fields 
formalism. Thus, consistent truncation 
is to demand that some components
of superfields possessing
ghost number different from zero
are defined to be vanishing
such that related antifields are also vanishing.

To discuss actions like (\ref{ef}) 
we need to introduce dual of a superfield.
In general $\Phi_D$ dual of a superfield
$\Phi$ in $n$ dimensions
is defined to satisfy
\be
\lb{dcg}
N_g (\Phi_D)=n-N_g(\Phi).
\ee
The components of $\Phi_D$ should be chosen such that 
they are related to
components of $\Phi$   with the 
correct ghost number attribution
and they may also be
Hodge duals of them.
 The components which cannot
fulfill these conditions should be eliminated by setting to zero
as it will be clarified in examples.

We first study
some examples in  4 dimensions which reveal the
main properties of the truncation of superfields. 
We also present a formulation of on--shell
$\rm {N}=1$ supersymmetric Yang--Mills theory
by superfields, inspired by the observed relations. 
Then we will discuss some other dimensions which 
clarify the relation of the generalized fields
and superfields methods for  the usual gauge theories 
and also for topological quantum field theories.

\subsection{Examples in 4 dimensions }

In 4 dimensions the superfields 
$U^{(4)}$ and  $V^{(4)}$
should satisfy
\be
\lb{4gc}
N_g(U_A^{(4)})+N_g(V_A^{(4)})=3.
\ee
Therefore, considering only positive
ghost numbers and the condition (\ref{be}),
there are two  possible  choices:
\beqa
& {\rm I.}&  N_g\left( U^{(4)}_I\right)=1 , 
N_g\left( V_I^{(4)}\right)=2 , \lb{c1}\\
&{\rm II.} &    N_g\left(U^{(4)}_{II}\right)=0 ,  
N_g\left(V_{II}^{(4)}\right)=3. \lb{c2}
\eeqa
Moreover, in each case one should  choose 
one of the superfields 
to be Grassmann even and the other 
to be  Grassmann  odd. 
Though all fields are taking values in a Lie algebra, we suppress 
trace over group elements.

Let us deal with   case I 
 calling 
$U_I$ and $V_I$ as $a$ and $b$ which can be written in terms
of components 
depending only on $x$ as
\beqa
U_I^{(4)}\equiv & a(x,\tau )  
 = &  a_0+\tau_\mu a_1^\mu +\tau_\mu\tau_\nu a_2^{\mu \nu}
+\tau_\mu\tau_\nu \tau_\rho a_3^{\mu \nu \rho} 
+\tau_\mu\tau_\nu \tau_\rho \tau_\si a_4^{\mu \nu \rho \si } ,\\
V_I^{(4)}\equiv &
b(x,\tau )   = &  b_0+\tau_\mu b_1^\mu +\tau_\mu\tau_\nu b_2^{\mu \nu}
+\tau_\mu\tau_\nu \tau_\rho b_3^{\mu \nu \rho} 
+\tau_\mu\tau_\nu \tau_\rho \tau_\si b_4^{\mu \nu \rho \si } .
\eeqa
Ghost number and Grassmann parity of the  components are
\[
\begin{array}{rrrrrrrrrrr}
& a_0 & a_1 & a_2 & a_3 & a_4 ;\ 
& b_0 & b_1 & b_2 & b_3 & b_4 \\
N_g & 1 & 0 & -1 & -2 & -3 ;\
& 2 & 1 & 0 & -1 & -2 \\
\ep & 1 & 0 & 1 & 0 & 1 ;\
& 0 & 1 & 0 & 1 & 0
\end{array}
\]
The action is
\be
\lb{saa}
S(a,b)= -\int d^4xd^4\tau  \left( bDa + \Vc (a,b)\right).
\ee

Analogous to the form of the action in terms of
generalized fields (\ref{ef}) 
we would like to consider for
$\Vc$  the following choices:
\beqa
& i. & \Vc_1=\fr{1}{2} b_Db + \fr{1}{2}b[a,a] ,\\
& ii. & \Vc_2 =\fr{1}{2} a_Da + \fr{1}{2} b [a,a] ,
\eeqa
where $a_D$ and $b_D$ are  duals of $a$ and $b$  satisfying
$N_g (a_D)=3,\ N_g (b_D)=2.$ Commutator denotes 
antisymmetrization of the Lie algebra generators.  

When we deal with $\Vc_1$ a consistent truncation is to
keep $N_g=1$ component of $a$ and set $b_0=b_1=0.$ Thus
antifields of them should also be taken as
$a_3=a_4=0.$ Now, by renaming the field components
we write the truncated superfields 
\beqa
a_{t1} & = & \eta + \tau^\mu A_\mu +\fr{1}{2}\tau^\mu \tau^\nu 
B^\star_{\mu \nu} , \lb{gf1} \\
b_{t1}& = & \fr{1}{4}\tau^\mu \tau^\nu \ep_{\mu \nu \rho \si}B^{\rho  \si} +
\fr{1}{3!}\tau^\mu \tau^\nu \tau^\rho\ep_{\mu \nu \rho \si} A^{\star \si} 
+\fr{1}{4!}\tau^\mu \tau^\nu \tau^\rho \tau^\si 
\ep_{\mu\nu\rho\si}\eta^\star ,\lb{gf2}
\eeqa
where as usual $\star$ indicates antifields.
The dual superfield can be written as
\be
\lb{bt1d}
b_{t1D} =  \fr{1}{2}\tau^\mu \tau^\nu B_{\mu \nu } +
\fr{1}{3!}\tau^\mu \tau^\nu \tau^\rho\ep_{\mu \nu \rho \si} A^{\star \si} 
+\fr{1}{4!}\tau^\mu \tau^\nu \tau^\rho \tau^\si 
\ep_{\mu\nu\rho\si}\eta^\star .
\ee

Substituting the superfields $a,\ b,\ b_D$ with the
truncated fields (\ref{gf1}--\ref{bt1d}) in (\ref{saa})
with $\Vc_1$ one obtains
\beqa
S_{4,YM} & \equiv & -\int d^4xd^4\tau   [ b_{t1}Da_{t1} + \Vc_1(a_{t1},b_{t1}) ]=
-\int d^4x\ (\fr{1}{2}B_{\mu \nu}F^{\mu \nu}  \nonumber \\
& & -B^{\mu\nu} \eta  B_{\mu\nu}^\star 
+A_\mu^\star D^\mu \eta +
\fr{1}{2}\eta^\star [ \eta , \eta ] 
-\fr{1}{4} B_{\mu\nu} B^{\mu\nu}),
\eeqa
where $F_{\mu \nu}=\del_\mu A_\nu -\del_\nu A_\mu + [A_\mu,A_\nu]$
and $D_\mu \eta  =\del_\mu \eta  + [A_\mu ,\eta].$ 

We may perform a partial gauge fixing $B^\star =0$ and then
use the equations of motion with respect to $B_{\mu\nu}$ to obtain
\[
S_{4,YM}=- \int d^4x\ (\fr{1}{4}F_{\mu \nu}  F^{\mu \nu}
+A^\star_\mu    D^\mu \eta
+\fr{1}{2}\eta^\star [ \eta ,\eta ] ),
\]
which is the minimal solution of the master equation for
Yang-Mills theory. Indeed, the components of the
fields (\ref{gf1}) and (\ref{gf2}) are the same with
the components of the generalized fields given in \cite{gf1}.

For $\Vc_2$ there is another consistent truncation: $a_0=0$
and its antifield $b_4=0.$
We rename the components to write the
truncated superfields as
\beqa
a_{t2} & = & \tau^\mu  A_\mu + \fr{1}{2}\tau^\mu \tau^\nu 
B^\star_{\mu \nu} 
+\fr{1}{3!} \tau^\mu \tau^\nu \tau^\rho \ep_{\mu \nu \rho \si}
C_0^{\star \si}
+ +\fr{1}{4!}
\tau^\mu \tau^\nu \tau^\rho \tau^\si \ep_{\mu \nu \rho \si}
C_1^\star , \lb{at2}\\
b_{t2} & = & C_1 +\tau^\mu C_{0\mu} 
+\fr{1}{4}\tau^\mu \tau^\nu \ep_{\mu \nu \rho \si} B^{\rho \si} 
+\fr{1}{3!} \tau^\mu \tau^\nu \tau^\rho 
\ep_{\mu \nu \rho \si }A^{\star \si}. \lb{bt2}
\eeqa

In general the dual superfield $a_D$ possesses 
components satisfying
\[
\begin{array}{rrrrrr}
& a_{D0} & a_{D1} & a_{D2} & a_{D3} & a_{D4} \\ 
N_g & 3 & 2 & 1 & 0 & -1 \\
\ep & 1 & 0 & 1 & 0 & 1 
\end{array}
\]
Components of $a_D$  should be chosen such that 
they are related to
components of $a$ or their Hodge duals
with   
correct ghost number. The components which cannot
fulfill these conditions should be defined  to vanish.
Thus, for the truncation (\ref{at2})
the only non-vanishing component is $a_{t2D3}$. Indeed,
we define
\[
a_{t2D}=\fr{1}{3!}\tau^\mu\tau^\nu\tau^\rho \ep_{\mu \nu \rho \si} A^\si.
\]
Thus, one obtains 
\beqa
S_{4,AS} & \equiv & -\int d^4xd^4\tau [b_{t2}Da_{t2} 
+\Vc_2(a_{t2},b_{t2})]=
-\int d^4x\ \{ \fr{1}{2}B_{\mu\nu}F^{\mu\nu} 
  \nonumber \\
& & +\fr{1}{2}
\ep_{\mu \nu \rho \si }C_0^\mu D^\nu B^{\star \rho \si}
 +C_1 D^\mu C^\star_{0\mu} +\ep^{\mu \nu \rho \si }
C_1 B^\star_{\mu \nu } B_{\rho \si }^\star  -
\fr{1}{4}A_\mu A^\mu \} .
\eeqa
This is the BV--master action of the self dual 
antisymmetric tensor field which was obtained 
in terms of the generalized fields whose components
coincide with the components of the superfields
(\ref{at2}, 
\ref{bt2}).

Let us deal with case  II (\ref{c2}).
Let us denote  the fields 
$U_{II}$ 
 and $V_{II}$ as $\Psi^\al $ and $\bar{\Psi}_\al ,$
where $\al$ is an index which will be specified below
and supposed to be lowered  with
an appropriate metric.
Ghost numbers of the components of $\Psi$ and $\bar{\Psi}$ 
follow from the definition (\ref{c2}) and their 
Grassmann parity are chosen as
\[
\begin{array}{rrrrrrrrrrrr}
& \Psi_0 &  \Psi_1 &  \Psi_2 &  \Psi_3 &  \Psi_4; &&
\bar{ \Psi}_0 &  \bar{\Psi}_1 &  \bar{\Psi}_2 & 
 \bar{\Psi}_3 &  \bar{\Psi}_4  \\ 
N_g & 0 & -1 & -2 & -3 & -4; &&
3 & 2 & 1 & 0 & -1 \\
\ep & 1 & 0 & 1 & 0 & 1; &&
0 & 1 & 0 & 1 & 0
\end{array}
\]
A consistent truncation is to set all the ghost fields
(positive ghost number carrying fields)
equal to zero, i.e. no gauge invariance
\beqa
\Psi_{t\al} & = & \Psi_{0\al}+\tau_\mu \Psi_{1\al}^\mu , \lb{tp1}\\
\bar{\Psi}_{t\al} & = & \tau_\mu \tau_\nu \tau_\rho 
\bar{\Psi}_{3\al}^{\mu \nu \rho}
+ \tau_\mu \tau_\nu \tau_\rho \tau_\si 
\bar{\Psi}_{4\al}^{\mu \nu \rho \si}. \lb{tp2}
\eeqa
We define
\beqa
\bar{\Psi}_{3\al}^{\mu \nu \rho} & =  &
\fr{1}{3!}
\ep^{\mu \nu \rho \si} \bar{\Psi}_{\al\si} \\
\bar{\Psi}_{4\al}^{\mu \nu \rho \si} & = & 
\fr{1}{4!}
\ep^{\mu \nu \rho \si} \Psi^\star_{0\al}, \\
\Psi_{1\al}^\mu & = &\bar{\Psi}_\al^{\star \mu},
\eeqa
where star indicates  antifields as usual.
By substituting the superfields
(\ref{c2})  with  the truncated superfields 
(\ref{tp1}, \ref{tp2}) in the action (\ref{sma}) one finds
its kinetic part  as
\be
\lb{spsi}
S_\Psi=\int d^4x d^4\tau \bar{\Psi}_{t\al} D \Psi_{t}^{\al}
= \int d^4x \bar{\Psi}_{\al \mu} \del^\mu \Psi_0^\al .
\ee
Now, let $\al$ be the spinor index and 
\be
\bar{\Psi}_\al^\mu = \fr{1}{2}\bar{\Psi}_{0\beta} 
\gamma^{\mu \beta}_{\al} ,
\Psi_{1\mu}^\al = \gamma_{\mu\beta }^\al  \bar{\Psi}^{\star\beta }_0  .
\ee
Thus, the action (\ref{spsi}) becomes
\be
S_\Psi =-\fr{1}{2}\int d^4x \bar{\Psi}_{0\al} 
\gamma^{\al}_{\mu \beta} \del^\mu 
\Psi_{0}^{\beta} .
\ee

Obviously, one can also deal with a theory which 
is a mixture of case I and II. Thus,
we can couple this theory to the Yang--Mills theory by
\[
S_{int}\equiv   \int d^4x d^4 \bar{\Psi}_{t\al}
[a_{t1} , \Psi_t^\al] = - \int d^4x \left[
 \fr{1}{2}\bar{\Psi}_0\gamma^\mu [A_\mu , \Psi_0 ]
- [ \bar{\Psi}_{0} ,\eta ] \bar{\Psi}_0^\star
- {\Psi}_{0}^\star [\eta , \Psi_0 ] \right].
\]
Thus the 4--dimensional
$N=1$ on--shell
supersymmetric Yang--Mills
 theory can be written in
terms of the superfields as
\be
S_{SYM}=S_{4YM}+S_\Psi+S_{int} .
\ee

Supersymmetric case was not discussed in terms of 
generalized fields method.
Here it follows as a natural consequence of truncating
superfields to obtain generalized fields.

Truncations which we studied do not  exhaust 
all of the possibilities. We studied examples
which are illustrating the procedure of truncation
for some basic  gauge theories.

\subsection{Examples in other dimensions}

Equipped with the detailed knowledge
of how generalized fields and superfields approaches
are related for 
4--dimensional theories
we can   discuss
their relation in other dimensions without
focusing on the  details.

In one dimension we have shown that generalized fields
method leads to BV--formulation of first order Lagrangians
for some constrained systems\cite{gf1}. 
By using the formulation  given 
in \cite{gd}  for phase space variables one can easily
observe that both methods lead to the same conclusions.
To clarify it let us deal with the superfields
\beqa
Q_\mu =q_\mu +\tau q_{1\mu}, &
P_\mu =p_\mu +\tau p_{1\mu}, & \ep (Q) =\ep (P) =0; \nonumber \\
E =e_0 +\tau e_1 ,&
F =f_0 +\tau f_1, & \ep(E)=\ep(F)=1,   \nonumber
\eeqa
where $\mu=1,\cdots,d.$ These 
will be shown to be suitable for 
the relativistic particle.
Let us attribute the ghost numbers 
\beqa
N_g(Q)=0, &
N_g(P)=0, \\
N_g(E)=1, &
N_g(F)=-1.
\eeqa
We permit a negative ghost number because
we need a Lagrange multiplier.
Let us deal with the action 
\be
\lb{sp}
S_p=\int dt d\tau [P_\mu D Q^\mu +EDF-\fr{1}{2}E P_\mu P^\mu ].
\ee
By renaming the components as
\[
q_{1\mu}=-p^\star_\mu,\
p_{1\mu}=q^\star_\mu ,\
e_0=\eta ,\
e_1=e ,\
f_0=e^\star ,\
f_1=\eta^\star ,
\]
where $\eta $ is ghost field and as usual star indicates antifields,
the action (\ref{sp}) reads
\[
S=\int dt\  [p\cdot \fr{\del q}{\del t} +e^\star \fr{\del \eta}{\del t}
 -\fr{1}{2}
e p^2 + q^\star \cdot p \eta ],
\]
which is the minimal solution of the BV--master equation
for the relativistic particle. 
Let us clarify relation between the two approaches.
In generalized fields method
the fields are grouped as
\beqa
\ti{q} & =  &  q^\mu_{(1+0+0)}+e_{(0+1+0)}  +\eta_{(0+0+1)}-
p^{\star \mu}_{(1+1-1)}, \lb{qti}  \\
\ti{p} & =  &  p_{\mu (d-1+0+0)} +q^\star_{\mu (d-1+1-1)}
+e^\star_{(d+0-1)} + \eta^\star_{(d+1-2)}, \lb{pti}
\eeqa
where the numbers in the parenthesis
indicate, respectively, grading due to space-time,
grading due to 1--dimensional manifold and ghost number.
The relevant action is
\be
S_p=\int dt\left[ \ti{p}\fr{\del \ti{q}}{\del t} 
+ \fr{1}{2} \ti{q} \ti{p} \ti{p}\right] .
\ee
Here multiplication is defined such that all three of the
gradings of  the action
$S_p$ vanish. This property leads to
the fact that  indeed the generalized fields (\ref{qti}, \ref{pti})
can be imagined as composed of two objects carrying
different indices.
Thus, the field contents of  both methods are the same.
 
In 2 dimensions  generally one deals with  theories where 
there is no need of any truncation to obtain the similar
results in either generalized or superfield formalism.
The superfields $U^{(2)}$ and $V^{(2)}$
can only have the ghost numbers
$N_g(V^{(2)}) =0,\ N_g(U^{(2)}) =1,$ 
when we restrict the ghost number to be positive.
We choose  $U^{(2)}$ to be Grassmann odd
and $V^{(2)}$ to be Grassmann even. Now,
each superfield has three components
\footnote{Here and in the following fields are Lie 
algebra valued but we suppress traces.}:
\beqa
U^{(2)} \equiv u & = 
& u_0+\tau_\mu u_1^\mu +\tau_\mu \tau_\nu u_2^{\mu \nu}, \lb{2d1}\\
V^{(2)} \equiv v & = 
& v_0+\tau_\mu v_1^\mu +\tau_\mu \tau_\nu v_2^{\mu \nu}. \lb{2d2}
\eeqa 
We deal with the action 
\be
\lb{2ym}
S_{(2)}= -\int d^2xd^2\tau\  (v Du +\fr{1}{2}v_D v + \fr{1}{2}v[u,u]) ,
\ee
where due to ghost number constraint $N_g(v_D)=2,$
two of the  components of 
$v_D$  should vanish and we write
\be
\lb{2dd}
v_D=\tau_\mu \tau_\nu \ep^{\mu \nu}v_0 .
\ee
Observe that one obtains the BV--master 
action of 2--dimensional
Yang--Mills theory in first
order formalism  
when we use (\ref{2d1}, \ref{2d2}, \ref{2dd})
in the action $S_{(2)}.$ In generalized fields
method one has the same field content with the 
action in the form of (\ref{2ym}).

When ghost numbers of superfields are restricted to be positive,
in 3 dimensions we have two possibilities:
\beqa
& 1.&\ N_g(U_1^{(3)})=1, N_g(V_1^{(3)})=1, \\
& 2.&\ N_g(U_2^{(3)})=0, N_g(V_2^{(3)})=2.
\eeqa
Let us  briefly discuss these cases.
In  case 1 we can take 
both of the fields to be the same and
Grassmann odd: $U_c.$  
By taking the action 
$$S_c=\int d^3x d^3\tau (U_cDU_c +
\fr{1}{3}U_c[U_c,U_c] ) $$
the BV--master
action of Chern-Simons theory
follows\cite{bm}. In fact generalized fields method
yields the same field content and
the same form of the action\cite{gf2}.
However, we can take two  different fields  $U_1$ and $V_1$
possessing the same ghost number. In this case 
with the appropriate action 
$$S_{3YM}=\int d^3xd^3\tau
(V_1DU_1 +\fr{1}{2}V_{1D}V_1 + \fr{1}{2} V_1[U_1,U_1]),$$
3--dimensional Yang--Mills
theory can be found by truncating the fields
in accordance with 4--dimensional formalism.
Case 2 can also be treated similar to 4--dimensional
case in terms of spinor fields with an appropriate 
truncation and action.

In  5 dimensions we have the choices:
\[
\begin{array}{lcc}
& N_g(V^{(5)}) & N_g(U^{(5)}) \\
(i). & 2  &  2 \\
(ii). & 3  &  1 \\
(iii). & 4  &  0 
\end{array}
\]  
for positive ghost numbers.
Let us discuss some possibilities.
Case $(i)$ can be discussed by
setting $U_1^{(5)}\equiv V_1^{(5)}=U$ and 
taking the action 
$$S_1=\int d^5x d^5\tau (UDU + \fr{1}{3}U[U,U]) .$$
An example to this case is studied below.
Case $(ii)$  with the action
$$S_2=\int d^5x d^5\tau \left( V_2^{(5)}DU_2^{(5)} 
+\fr{1}{2}V_{2D}^{(5)}V_2^{(5)} 
+ \fr{1}{2} V_2^{(5)}[U_2^{(5)},U_2^{(5)}]\right) $$
yields Yang--Mills theory in 5 dimensions after
an appropriate truncation of ghost fields
according to its general fields formulation.
 Similarly case
$(iii)$ can be written 
in terms of spinor fields with
appropriate action 
and truncation of superfields. 
Obviously, by some other choices of actions
and truncations one can obtain different theories.

We would like to discuss a theory which elucidates the essential
points of the relation between the two methods for
topological quantum field theories 
 which is also
an example to case $(i).$
There exists a generalized Chern--Simons theory\cite{mp}
 in any odd
dimension  $d=2n+1,$ which was considered in terms of 
generalized fields method\cite{gf2}
yielding the BV--master action 
\be
\lb{sd}
S_d=\fr{1}{2}\int  _{M_d} \left( \ti{A} d \ti{A}
+ \fr{2}{3} \ti{A}^3 \right) ,
\ee
where
 $\ti{A} =\ti{\phi}+\ti{\psi}$ defined as\footnote{There are 
some typos in \cite{gf2} whose corrected versions are
given in (\ref{ft}).}
\beqa
\ti{\phi}  &  =  &
\sum_{i=0}^{n-1}  \left[ \phi_{(2i+1,0)}
+ \sum_{j=1}^{2i+1} \eta_{(2i+1-j,j)}  +\phi^*_{(2i+2,-1)}
+ \sum_{j=-2n+2i}^{-2} \eta^*_{(2i+1-j,j)} \right]  \nonumber \\
 & &  \lb{ft} \\
\ti{\psi} &  =  &
\sum_{i=0}^n  \psi_{(2i,0)}
+\sum_{i=1}^n \sum_{j=1}^{2i} \kappa_{(2i-j,j)}
+\sum_{i=0}^n \psi^*_{(2i+1,-1)}
+ \sum_{i=0}^{n-1} \sum_{j=-2n+2i-1}^{-2} \kappa^*_{(2i-j,j)}  .\nonumber
\eeqa
The antifield of the field $a_{(k,l)}$ is defined as $a^*_{(2n+1-k,-l-1)}$.
In terms of
$\ti{\phi}$ and $\ti{\psi}$ one can write 
(\ref{sd}) as
\be
\lb{fa}
S_d =\fr{1}{2} \int  _{M_d} \left( \ti{\phi} d \ti{\phi} 
+\fr{1}{3}\ti{\phi}[\ti{\phi},\ti{\phi}]
+\ti{\psi}d\ti{\psi}+\ti{\psi} [\ti{\phi}, \ti{\psi}] \right) .
\ee

Let us discuss this theory
in 5 dimensions to
illustrate
how generalized fields and superfields are related
for topological quantum field theories. 
In terms of superfields whose
ghost numbers and Grassmann parities are
\[
\begin{array}{cccccc}
    & \Phi_1 & \Phi_3 & \Psi_0 & \Psi_2 & \Psi_4 \\
N_g & 1      &      3 &      0 &      2 & 4 \\
\ep & 0      &      0 &      1 &      1 & 1 
\end{array}
\]
one can write the action
\beqa
S_{s5} & = & \int_{M_5} d^5\tau [
\Phi_1D\Phi_3 +\Psi_0D\Psi_4+\fr{1}{2}\Psi_2D\Psi_2 
+\fr{1}{3} 
 \Phi_3 [\Phi_1 ,\Phi_1] 
+ \Psi_0 [ \Phi_1 ,\Psi_4 ]
\nonumber \\
&& + \Psi_0 [\Phi_3 ,\Psi_2] 
+\fr{1}{2} \Psi_2 [\Phi_1 ,\Psi_2]
 ] , \lb{5dt}
\eeqa

Components of these superfields and the components
of  generalized fields (\ref{ft}) are in one to one
correspondence. Indeed, superfields
can be written as
\beqa
\Phi_1 & = & \eta_{(0,1)}+ \tau \phi_{(1,0)}
+\tau^2\phi^\star_{(2,-1)}
+\tau^3\eta^\star_{(3,-2)}
+\tau^4\eta^\star_{(4,-3)}
+\tau^5\eta^\star_{(5,-4)}, \\
\Phi_3 & = & \eta_{(0,3)}+ \tau \eta_{(1,2)}
+\tau^2\eta_{(2,1)}
+\tau^3\phi_{(3,0)}
+\tau^4\phi^\star_{(4,-1)}
+\tau^5\eta^\star_{(5,-2)} ,\\
\Psi_0 & = & \psi_{(0,0)}
+\tau^1\psi^\star_{(1,-1)}
+\tau^2\ka^\star_{(2,-2)}
+\tau^3\ka^\star_{(3,-3)}
+\tau^4\ka^\star_{(4,-4)}
+\tau^5\ka^\star_{(5,-5)} ,\\
\Psi_2 & = & \ka_{(0,2)}
+\tau^1\ka_{(1,1)}
+\tau^2\psi_{(2,0)}
+\tau^3\psi^\star_{(3,-1)}
+\tau^4\ka^\star_{(4,-2)}
+\tau^5\ka^\star_{(5,-3)} ,\\
\Psi_4 & = & \ka_{(0,4)}
+\tau^1\ka_{(1,3)}
+\tau^2\ka_{(2,2)}
+\tau^3\ka_{(3,1)}
+\tau^4\psi_{(4,0)}
+\tau^5\psi^\star_{(5,-1)} ,
\eeqa
where a shorthand notation is used for the odd coordinates
$\tau^i\equiv \tau_{\mu_1}\cdots \tau_{\mu_i}.$
One can observe that by these definitions 
(\ref{fa}) in  5 dimensions  and (\ref{5dt})
coincide:
\[
S_5=S_{s5}.
\]
Higher dimensions can be studied in a similar manner.

\section{Discussions}

Generalized fields and
superfields formulation
of finding solutions of the BV--master equation are shown
to be related  by 
consistent truncations of the latter
for the usual gauge theories. For topological quantum field
theories they yield the same field contents
without any truncation. Though the latter relation is shown
for specific cases, we believe that it is a general conclusion.
 Generalized fields method is
based on  a classical
 gauge theory action. However, 
superfields approach 
begin with a BV--master action from which
one can read the underlying  classical gauge
theory by eliminating
ghost variables
and antifields.
Having a connection between these theories we can formulate
different kind of gauge theories in terms  of superfields
as it is illustrated for supersymmetric Yang--Mills theories.
Moreover, this  relation can give some hints  to discover  
some other  theories which can be discussed  
either in terms of generalized fields or superfield algorithms
and suitable to describe some physical systems.

\end{document}